%% file: jcgipa-rev.tex
\documentclass[aps,floatfix,pra,reprint,superscriptaddress]{revtex4-2}

\usepackage{amsmath}
\usepackage{amssymb}
\usepackage{amsthm}
\usepackage{bbm}
\usepackage{color}
\usepackage{float}
\usepackage{graphicx}
\usepackage{hyperref}
\usepackage{physics}
\usepackage[T1]{fontenc}
\usepackage{times}
\usepackage{txfonts}
\usepackage[utf8]{inputenc}
\usepackage{xcolor}
\usepackage[normalem]{ulem}
\usepackage{lineno}

\hypersetup{
    colorlinks=true,linkcolor=blue,citecolor=blue,
    filecolor=blue,urlcolor=blue,breaklinks=true
}

\newcommand{\orcid}[1]{\href{https://orcid.org/#1}
{\includegraphics[width=7pt]{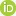}}}

\theoremstyle{definition}

\def\be{\begin{equation}}
\def\ee{\end{equation}}

\def\bc{\begin{center}}
\def\ec{\end{center}}
\def\bal{\begin{align}}
\def\eal{\end{align}}

\begin{document}

\title{
  Reproducing the effects of quantum deformation in the undeformed
  Jaynes-Cummings model
}

\author{Thiago T. Tsutsui\orcid{0009-0001-1654-0330}}
\email{takajitsutsui@gmail.com}
\affiliation{
 Programa de Pós-Graduação em Ciências/Física,
 Universidade Estadual de Ponta Grossa,
 84030-900 Ponta Grossa, Paraná, Brazil
}
\author{Antonio S. M. de Castro\orcid{0000-0002-1521-9342}}
\email{asmcastro@uepg.br}
\affiliation{
 Programa de Pós-Graduação em Ciências/Física,
 Universidade Estadual de Ponta Grossa,
 84030-900 Ponta Grossa, Paraná, Brazil
}
\affiliation{
  Departamento de Física,
  Universidade Estadual de Ponta Grossa,
  84030-900 Ponta Grossa, Paraná, Brazil
}

\author{Fabiano M. Andrade\orcid{0000-0001-5383-6168}}
\email{fmandrade@uepg.br}
\affiliation{
  Programa de Pós-Graduação em Ciências/Física,
  Universidade Estadual de Ponta Grossa,
  84030-900 Ponta Grossa, Paraná, Brazil
}
\affiliation{
  Departamento de Matemática e Estatística,
  Universidade Estadual de Ponta Grossa,
  84030-900 Ponta Grossa, PR, Brasil
}

\date{\today}

\begin{abstract}
In the Jaynes-Cummings (JC) model, the time dependence in the coupling parameter allows changes in the forms of the Rabi oscillations. In the inverse problem approach (IPA), the time-dependent coupling parameter is obtained from the resulting population inversion. In this work, we employ the IPA to obtain a time-dependent coupling that reproduces the effects of $\kappa$-deformation in the population inversion of an undeformed JC. This is relevant because it may pave the way for simulating quantum deformation in the JC model and possibly enable an experimental verification in a setting where the coupling can be precisely controlled.
\\

\noindent \textbf{Adv Quantum Technol. 2400559 (2025)}; doi: %
\href{https://doi.org/10.1002/qute.202400559}
{10.1002/qute.202400559}
\end{abstract}

\maketitle

\section{Introduction}
\label{sec:introduction}
The Jaynes-Cummings (JC) model \cite{SHORE1993,GERRY2005,LARSON2021} is
well-established in quantum optics.
It was introduced in 1963 \cite{JAYNES1963} as a theoretical model that
describes the quantum dynamics of a two-level atom interacting with a
mode in a quantum cavity.
The model highlights relevant aspects such as the Rabi oscillations
occuring due to the interaction between matter and quantized radiation.
In 1987, the model underwent experimental validation via a one-atom
maser \cite{REMPE1987}, wherein the collapse and revival of the Rabi
oscillations within the atomic population were observed.
The list of applications of the JC model is vast, addressing different
scenarios in relativistic systems
\cite{Rozmej1999,BERMUDEZ2007,Bermudez2008}, quantum information
\cite{Romanelli2009,Meher2022}, quantum computation \cite{Monroe1995},
and implementations in trapped ions
\cite{Blockley1992,Vogel1995,Pedernales2015,
  Chaichian1990,Santos2012,Dehghani2016}.

The time-dependent JC model \cite{LARSON2021,DeCastro2023} is an
extension of the original framework, where the parameters, otherwise
considered constant, evolve over time.
Naturally, this generalization allows for changes in the Rabi
oscillations.
In this context, we focus on the scenario of a time-dependent coupling
parameter under the resonance condition
\cite{Schlicher1989,Larson2003a,Larson2003b}.
In Ref. \cite{Yang2006}, Yang \emph{et al.} proposed an inverse problem
approach (IPA), in which the time-dependent coupling parameter is
derived directly from the resulting population inversion with a formula,
considering resonance. 
Using this method, we can explore the mimicry of physical scenarios
within the JC model through their effects on population inversion,
considering a time-dependent coupling.

An interesting approach to studying the JC model is the application of
deformed algebras to create alternative versions of the model
\cite{Chaichian1990,Santos2012,Dehghani2016}, introducing additional
noncommutativiy in the algebraic structures, and with significant
changes in the system dynamics.
Particularly intriguing is the use of the $\kappa$-deformed
Poincaré-Hopf algebra \cite{Lukierski1991,Lukierski1992,Lukierski1994},
where $\kappa$ represents a fundamental deformation parameter.
Since the Poincaré-Hopf algebra is generally associated with a flat
spacetime and the parameter $\kappa$ has mass dimension, this scenario
is usually  related to quantum gravity \cite{Freidel2006,Lukierski2017}.
In this context, Uhdre \emph{et al.} \cite{UHDRE2022} presented the $\kappa$-Jaynes-Cummings ($\kappa$-JC) model.

In this work, we explicitly present the population inversion of the
$\kappa$-JC and obtain a coupling that reproduces the effects of
$\kappa$-deformation in the undeformed JC, employing the IPA.
This fact is relevant because it may lead to simulation of the quantum
deformation in the JC model and possibly enables an
unprecedented experimental verification in a setting where
the coupling can be precisely adjusted.

We organize the work as follows.
In Sec. \ref{sec:inverse_problem_approach}, we provide an overview of
the IPA and show how a particular coupling parameter can induce
collapses and revivals of Rabi oscillations in a cavity that is not
initially coherent.
Subsequently, we explicitly outline the $\kappa$-JC population inversion
and use the IPA to derive the associated coupling parameter in Sec.
\ref{kappa-JCM}.
In the sections above, we identify the coupling parameters that lead to
the required population inversion assuming an initial \emph{vacuum}
state for the cavity mode, a limitation stemming from how the IPA was
presented in the original reference \cite{Yang2006}.
Due to this, in Sec. \ref{sec:gipa}, we generalize the IPA to initial
cavity states described by a superposition and derive a coupling that
induces the population inversion from the $\kappa$-JC within an initial
coherent state in the cavity mode, allowing for a more straightforward
physical meaning of the connection between quantum deformation and
temporal dependence in the coup+ling parameter.
In this case, the coupling parameter serves as a link between the
time-dependent JC model and the $\kappa$-JC model.
Finally, in Sec. \ref{sec:conc} we present our conclusions.

\section{Inverse Problem Approach}
\label{sec:inverse_problem_approach}

In the rotating wave approximation (RWA), the JC Hamiltonian operator
reads (with $\hbar=1$)
\begin{equation}
  H = \frac{1}{2}\omega\sigma_{z}+\nu a^{\dagger}a +
  \lambda^{*} a\sigma_{+}+\lambda a^{\dagger}\sigma_{-},
  \label{eq:full_Hamiltonian_JC}
\end{equation}
in which $\omega$ and $\nu$ are the frequencies of the atom and cavity
mode, respectively, and $\lambda$ is the strength coupling parameter.
From the Pauli operators $\sigma_{k}$, $k=x,y,z$, we have
$\sigma_{\pm}=\sigma_{x}\pm i\sigma_{y}$ as the ladder operators,
generators of the  $\mathfrak{su}(2)$ algebra \cite{Klimov2009}, while
the cavity mode is represented by the annihilation ($a$) and creation
($a^{\dagger}$) operators, elements of the Weyl-Heisenberg algebra
\cite{Cantuba2024}.
The operators above obey the well-known commutation relations   $[\sigma_{+},\sigma_{-}] =  2\sigma_{z}$, and $[a,a^{\dagger}] = 1$.

From now on, we will incorporate temporal dependence into the coupling
by rewriting it as $\lambda(t)$.
To analyze the dynamics of the JC, we can use the interaction picture to
rewrite the  Hamiltonian, Eq. \eqref{eq:full_Hamiltonian_JC}, where
\begin{equation}
V(t)=\lambda^{*}(t)a\sigma_{+}+\lambda(t)a^{\dagger}\sigma_{-}, \label{eq:interaction_Hamiltonian_JC}
\end{equation}
which expresses how the interaction between the atom and the cavity
occurs under the resonance condition, i.e., when $\omega=\nu$.

Let us now consider the atom initially in the excited state
($\left|e\right\rangle$), and the field is in the vacuum state
($\left|0\right\rangle$), meaning that the initial state of the system
reads $\ket{\psi(0)}=\ket{e,0}$.
The dynamics represented by Eq. \eqref{eq:interaction_Hamiltonian_JC}
results in the initial state being coupled solely to the final state
$\ket{g,1}$.
Thus, for any time $t$, the system's state is described by
\begin{equation}
  \label{eq:state_JC}
  \ket{\Psi(t)} = C_{e}(t)\ket{e,0} + C_{g}(t)\ket{g,1},
\end{equation}
where $C_{e}(t)$ and $C_{g}(t)$ are the probability amplitudes
associated with atomic excited and ground states, respectively.
Naturally, they are associated with the states of the cavity mode,
something implicit for now.
For the case under investigation, the initial conditions agree with the
constraints  $|C_{e}(0)|^{2}=1$ and $|C_{g}(0)|^{2}=0$, and the
probability amplitudes must obey the normalization condition
$|C_{e}(t)|^{2}+|C_{g}(t)|^{2}=1$.
Furthermore, in describing the system dynamics, we focus our attention
on the experimentally measurable   \cite{LARSON2021,Haroche2006}
population inversion $W(t)$ defined as
\begin{equation}
  \label{eq:population_inversion}
  W(t)=\expval{\sigma_{z}} =
  \left|C_{e}(t)\right|^{2}-\left|C_{g}(t)\right|^{2}.
\end{equation}

For the time-independent JC with $\lambda(t)=\lambda_{0}$, and initial
state $\ket{e,0}$, the population inversion takes the simple form
$W(t)=\cos{(2\lambda_{0}t)}$ with $\lambda_0$ as a real number.
The periodic nature of $W(t)$ reflects the Rabi oscillations, with the
coupling influencing the rate at which they occur.
On the other hand, in the time-dependent JC model
\cite{LARSON2021,DeCastro2023}, the parameters of the Hamiltonian
\eqref{eq:full_Hamiltonian_JC} evolve with time.
This introduces an additional level of complexity to the system, as
analytical solutions are not always achievable.
In this work, we focus on the case where the time dependency is
incorporated into the coupling, considering resonance
\cite{Joshi1990,Prants1992,Joshi1993,Dasgupta1999}.
To the best of our knowledge, the first approach in this direction was
carried out by Schlicher \cite{Schlicher1989}, with a periodic coupling
representing atomic motion.
This variation of the original framework can introduce a new physical
aspect to the problem, such as transient effects in the cavity,
resulting in a population inversion that does not necessarily retain the
simple nature of the time-independent case.
In Ref. \cite{Maldonado-Mundo2012} was suggested that a time-dependent
coupling could be feasible in cavity quantum electrodynamics, with the
variation of the atom's position or the change in the field strength.

\begin{figure}[t]
  \centering
  \includegraphics[width=\columnwidth]{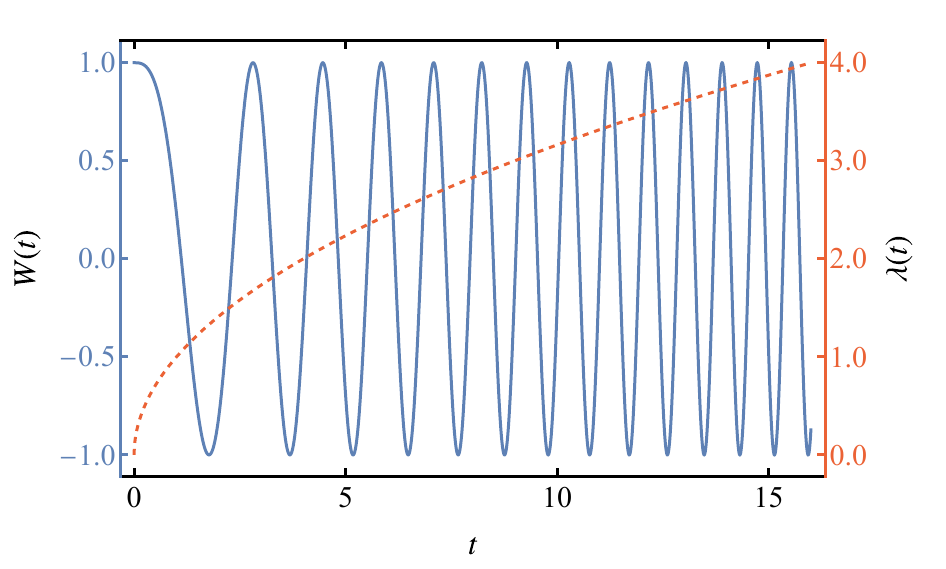}
  \caption{
    The population inversion $W(t)$ in Eq. \eqref{eq:W1} (blue, solid
    line) and the corresponding $\lambda(t)$ (orange, dashed line) with
    two different vertical scales for better visualization of the
    relationship between the coupling parameter and the Rabi
    oscillations.
  }
  \label{fig:fig1}
\end{figure}

Exploring various time-dependent scenarios, we consider different
functions of the dimensionless time scale $\zeta t$ for the coupling
strength $\lambda(t)$. 
For instance, if we employ $\lambda(t)=\lambda_{0}\sqrt{\zeta\,t}$, we
obtain the corresponding population inversion in the form
\begin{equation}
  W(t)=\cos{\left(\frac{4}{3}\, \lambda_{0} \sqrt{\zeta t^3} \right)}.
\label{eq:W1}
\end{equation}
This time-dependent coupling and the population inversion are
represented in Fig. \ref{fig:fig1}, assuming $\lambda_0=\zeta=1$, where
we can observe that increasing the coupling with time, faster are the
Rabi oscillations.

On the other hand, Yang \emph{et al.} proposed an IPA to investigate
different population inversions and their respective couplings in the
time-dependent JC \cite{Yang2006}.
Their method describes the time-dependent coupling as a function of the
required population inversion, when the initial state of the system is
$|e,0\rangle$. 
The equation that represents their method is
\begin{equation}
  \label{eq:coupling_factor_inverse_approach_ressonance}
  \lambda(t)= \frac  {-\dot{W}(t)} {2\sqrt{1-W^{2}(t)}}.
\end{equation}
Thus, the direct substitution of a specific form of $W(t)$ produces the
corresponding $\lambda(t)$, and by solving the system we can obtain the
coefficients $C_{e}(t)$ and $C_{g}(t)$ that results in the input
population inversion.
It is worth to mention that a similar modulation for the coupling could
be derived from the Bloch equations and applied for the JC model
\cite{SHORE1993}.
Rather than exploring other sources, we chose to base our work on the
referenced paper, as it uniquely addresses the issue we are examining:
achieving population inversion through tailored coupling.
As a trivial example, we can use the population inversion \eqref{eq:W1}
as input, obtaining $\lambda(t) = \lambda_{0} \sqrt{\zeta\,t}$.

An emblematic aspect of the JC is the collapses and revivals of the Rabi
oscillations when the initial state of the cavity mode is a coherent
state  $\ket{\alpha}$.
In this case, with the initial state of the system given by
$\ket{\Psi(0)} =\ket{e,\alpha}$, the population inversion reads
\cite{GERRY2005,LARSON2021}
\begin{equation}
  \label{eq:population_inversion_not_deformed_coherent}
  W_{\alpha}(t) = \sum_{n=0}^{\infty}\frac{\expval{n}^{n}
    e^{-\expval{n}}}{n!}\cos(\Omega_{n}t),
\end{equation}
with $\Omega_{n}=2\lambda_{0}\sqrt{n+1}$.
Here $\expval{n} =|\alpha|^2$ is the mean photon number corresponding to
the coherent state $\ket{\alpha}$.

\begin{figure}[t]
  \centering
  \includegraphics[width=1\linewidth]{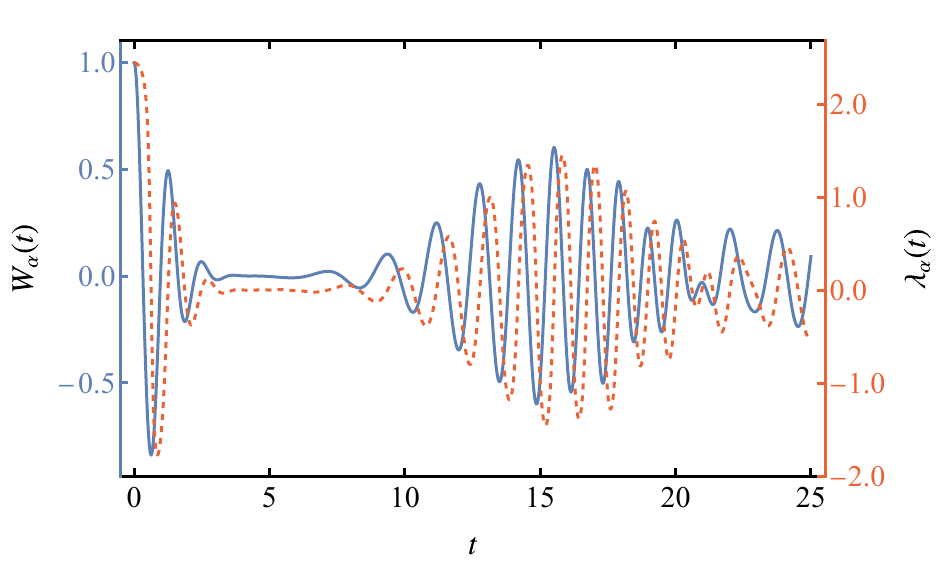}
  \caption{
    The population inversion $W_{\alpha}(t)$ of the initially coherent
    cavity in Eq.  \eqref{eq:population_inversion_not_deformed_coherent}
    (blue, solid) and the coupling parameter $\lambda_{\alpha}(t)$
    (orange, dashed) that  causes it in an initial vacuum state of the
    cavity, plotted employing $\lambda_0=1$ and $\expval{n}=5$.
    In this case, the coupling parameter reproduces the collapses and
    revivals of $W_{\alpha}(t)$.}
  \label{fig:fig2}
\end{figure}

Employing the IPA, we can reproduce the collapses and revivals assuming
the vacuum state on the cavity mode and the coupling parameter
$\lambda_{\alpha}(t)$ obtained from the substitution of Eq.
\eqref{eq:population_inversion_not_deformed_coherent} into Eq.
\eqref{eq:coupling_factor_inverse_approach_ressonance}.
In Fig.
\ref{fig:fig2}, we illustrate the population inversion $W_\alpha(t)$ and
the respective coupling $\lambda_{\alpha}(t)$ for
$\lambda_0=1$ and $\expval{n}=5$.
It is interesting to observe that the coupling reproduces the behavior of $W_{\alpha}(t)$ inheriting the behavior of the collapses and revivals.

\section{Quantum deformation in the JC}
\label{kappa-JCM}

The commutativity of the operators components, i.e.,
$\left[x_i,x_j\right]=\left[p_i,p_j\right]=0$, indicates that position
and momentum coordinates can be determined separately, reflecting the
presence of an underlying classical geometric structure, where spatial
and temporal coordinates commute, within Quantum Mechanics.
However, certain physical contexts, such as the presence of quantum
gravity effects, justify investigating situations where this
commutativity breaks down.
In general, the study of these scenarios is referred to as
\emph{quantum deformations} \cite{Lukierski2017}.

In quantum field theory, particles transform under unitary
representations of the Poincaré group, which encompasses symmetries of
translations, rotations, and changes of reference frames
\cite{Schwartz2014}.
The commutation relations arising from this group define the Poincaré
algebra.
The $\kappa$-Poincaré-Hopf algebra emerged in the 1990s
\cite{Lukierski1991,Lukierski1992,Lukierski1994} as a deformation of the
Poincaré algebra, with a deformation parameter symbolized by $\kappa$.
It is an important theory in the context of alternative scenarios for
the study of Quantum Mechanics, and has been applied to the
study the Landau problem \cite{Roy1994}, the Aharonov-Bohm spin-1/2
problem \cite{Andrade2013},  and the Dirac oscillator
\cite{Andrade2014a,Andrade2014b}.
The following commutation establishes this algebraic structure relations
\begin{align}
  \left[M_{i},M_{j}\right] = {}
  & \epsilon_{ijk}M_{k},\left[M_{i},P_{\mu}\right] =
    \left(1-\delta_{0\mu}\right)i\epsilon_{ijk}P_{k},
  \nonumber \\
  \left[L_{i},L_{j}\right] = {}
  & -i\epsilon_{ijk}\left[M_{k}\cosh\left(\varepsilon
    P_{0}\right)-\frac{\varepsilon^{2}}{4}P_{k}P_{l}M_{l}\right],
    \nonumber\\
  \left[P_{\mu},P_{\nu}\right] = {}
  & 0,  \quad  \left[M_{i},L_{j}\right]=i\epsilon_{ijk}L_{k},
    \nonumber\\
  \left[L_{i},P_{\mu}\right] = {}
  &  i\left[P_{i}\right]^{\delta_{0\mu}}  \left[\delta_{ij}\varepsilon^{-1}
    \sinh\left(\varepsilon P_{0}\right)\right]^{\left(1-\delta_{0\mu}\right)},
\end{align}
where $\varepsilon=1/\kappa$, $P_\mu$ are the $\kappa$-deformed
generators of energy and momenta with $M_i$ and $L_i$ corresponding to
the rotations and $\kappa$-deformed boost generators, respectively.
As the Poincaré algebra is associated with flat spacetime and $\kappa$
has dimensions of mass, it is believed that its deformed version is
related to quantum gravity \cite{Freidel2006,Lukierski2017}.
A short introduction to $\kappa$-deformations can be found in
Refs. \cite{Kowalski2017,Lukierski2017}.
Recently, Uhdre \emph{et al.} \cite{UHDRE2022} used the
$\kappa$-Poincaré-Hopf algebra to simultaneously present the
$\kappa$-deformed JC ($\kappa$-JC), and Anti-JC ($\kappa$-AJC) models,
utilizing a mapping introduced by Bermudez \emph{et al.}
\cite{BERMUDEZ2007}.
There, the authors determined the average values of various observables
for the $\kappa$-JC considering a time-independent coupling.
Interestingly, if the initial state on the cavity mode is one Fock
state, the average values of the observables in the deformed and
undeformed models showed no first-order corrections.
Furthermore, for a coherent initial state, a difference in the expected
values of the angular momentum arises.
For the scope of this work, the average value of spin in the
$z$-direction of the $\kappa$-JC considering a coherent initial state on
the cavity mode is of great interest since it is directly related to
population inversion.
Thus, assuming an initial coherent state of the cavity mode in the
$\kappa$-JC scenario and an excited state of the atom, the average value
of spin in the $z$-direction is given by Ref. \cite{UHDRE2022}, which is
given by
\begin{align}
  \label{eq:expected_spin}
  \expval{S_{z}(t)}_{\alpha}^{\epsilon} = {}
  &
    \frac{1}{2}-\sum_{n=0}^{\infty}\frac{\expval{n}^{n}e^{-\expval{n} }}{n!}
  \mathcal{S}_{n}(t)
    \nonumber \\
  &
    +\epsilon\sum_{n=0}^{\infty}\frac{\expval{n} ^{n+1}e^{-\expval{n}}}{n!}
    \left[\mathcal{S}_{n}(t)-\mathcal{S}_{n+2}(t)\right],
\end{align}
where $\epsilon=mc^{2}\varepsilon/2$ is a dimensionless deformation
parameter, and
\begin{equation}
  \label{eq:s_function}
  \mathcal{S}_{n}(t) = \frac{\Omega_{n}^2}  {\Delta^2 + \Omega_{n}^2}
  \sin^{2}\left[\frac{1}{2} \sqrt{\Delta^2+\Omega_{n}^2} t\right],
\end{equation}
with $\Delta= \omega-\nu$ meaning the detuning between atom and cavity
modes.
The way we present Eq. \eqref{eq:s_function} is different from Eq.
(52) of Ref. \cite{UHDRE2022} as we want to express it using an
exclusive JC notation.
The population inversion is twice the average value of the spin in the
$z$-direction in such a way that for the case of resonance ($\Delta=0$),
we have the population inversion for the $\kappa$-JC is given by
\begin{align}
  \label{eq:deformed_pop_inversion}
  W_{\alpha}^{\epsilon}(t)  = {}
  &
    \sum_{n=0}^{\infty}\frac{\expval{n} ^{n}e^{-\expval{n}}}{n!}
    \Biggl\{\cos(\Omega_{n})
    \nonumber\\
  &
    +\epsilon\expval{n}
    \left[
    \cos(\Omega_{n+2}t)-\cos(\Omega_{n})
    \right]
    \Biggr\},
\end{align}
which describes the dynamical behavior of the population inversion for
the $\kappa$-JC when the cavity mode is initially in a coherent state.
Note that for $\epsilon=0$, the population inversion recovers the
undeformed case, Eq. \eqref{eq:population_inversion_not_deformed_coherent}.
On the other hand, employing the IPA, we can reproduce the population
inversion for $\kappa$-JC employing a coupling parameter
$\lambda_\alpha^\epsilon(t)$, which is obtained from the substitution of
Eq. \eqref{eq:deformed_pop_inversion}, and assuming the cavity mode
initially in the vacuum state.
This is an interesting result since the factor $\epsilon$ becomes a
number in the coupling, and the effects of $\kappa$-deformation in the
population inversion can be reproduced by carefully adjusting it.
The possibility of emulating the effects of deformation through a
time-dependent coupling may pave the way for their observation.
Thus, we expect the reproduction of the deformation effects through a
time-dependent coupling may lead to new evidence of deformation in
experimental contexts \cite{LIU2018}.

\begin{figure}[t]
  \centering
  \includegraphics[width=\columnwidth]{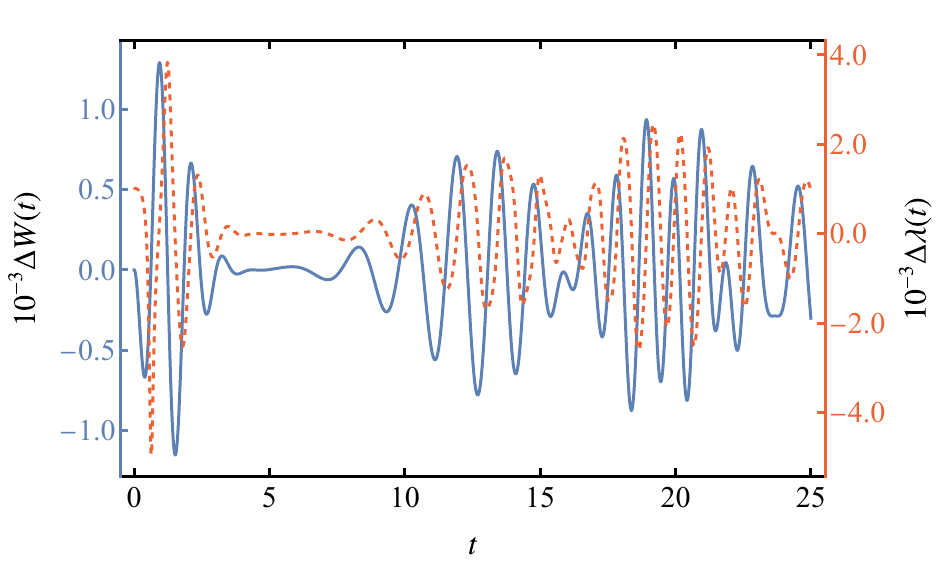}
  \caption{
    Difference between the population inversion $\Delta W(t)$ (blue,
    solid), and coupling parameter $\Delta \lambda(t)$ (orange, dashed)
    between the $\kappa$-deformed and undeformed JC models, using
    $\epsilon=5\times10^{-4}$, $\expval{n} = 5$ and $\lambda_0=1$.
  }
  \label{fig:fig3}
\end{figure}

We can observe the effects of the deformation on the behavior of the
population inversion by comparing the results for the JC and
$\kappa$-JC.
For this intent, let us define
\begin{align}
  \Delta W(t) = {}
  &  W_{\alpha}^{\epsilon}(t)-W_{\alpha}(t),
  \label{eq:D1} \\
  \Delta \lambda(t) = {}
  & \lambda_{\alpha}^{\epsilon}(t)-\lambda_{\alpha}(t),
  \label{eq:D2}
\end{align}
which measures the differences between the $\kappa$-deformed and
undeformed population inversion and the coupling parameter,
respectively.
In Fig. \ref{fig:fig3}, using $\epsilon=5\times10^{-4}$, which is an
upper bound for the deformation parameter \cite{Andrade2014a,UHDRE2022},
we show these differences.
We note that in the intervals where there is a difference between the
coupling parameter also there is a difference in the population
inversion.
So, differences in the coupling parameter lead to differences in the
population inversion.

\section{Generalized Inverse Problem Approach}
\label{sec:gipa}

In the previous section, we showed that it is possible to replicate the
population inversion $W^{\epsilon}_{\alpha}(t)$ in the  undeformed JC
by employing the time-dependent coupling
$\lambda^{\epsilon}_{\alpha}(t)$ when the cavity is initially in the
vacuum state.
However, it is of interest to seek a time-dependent coupling that causes
the same population inversion in the JC, but starting with an initial
coherent state in the cavity mode.
With this in mind, we establish a connection between the undeformed
time-dependent JC model and the $\kappa$-JC model by means of a
time-dependent coupling using the same initial state in both models.
To achieve this goal, we first need to generalize the IPA to a situation
encompassing more general states in the cavity.

\begin{figure}
  \centering
  \includegraphics[width=\columnwidth]{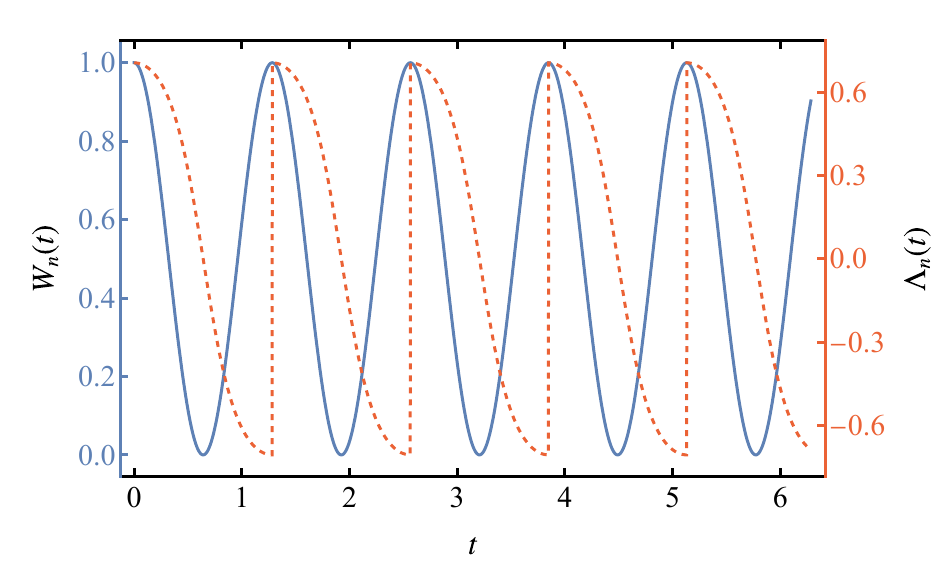}
  \caption{
    The population inversion $W_{n}(t)$ in
    Eq. \eqref{eq:pop_inversion_cos_squared} (blue, solid) and the
    coupling parameter $\Lambda_{n}(t)$ (orange, dashed) that causes it
    in an initial number state of the cavity mode, plotted employing
    $\lambda_0=1$ and $n=5$.
  }
  \label{fig:fig4}
\end{figure}

We start considering an initial state $\ket{\Psi(0)}=\ket{e,n}$ in the
JC system under the resonance condition.
Due to the algebraic structure of the JC model, the extension of Eq.
\eqref{eq:coupling_factor_inverse_approach_ressonance} to an initial
state with $n$ photons in the cavity, instead of the vacuum state, is
given by the rescaling factor $(n+1)^{-1/2}$.
In this manner, the equation governing the coupling parameter is given
by
\begin{equation}
  \label{eq:generalized_inverse_problem_coupling}
  \Lambda_n(t)=
  \frac{1}{2\sqrt{n+1}}\frac{-\dot{W_{n}}(t)}{\sqrt{1-W_{n}^{2}(t)}},
\end{equation}
which reduces to Eq. \eqref{eq:expected_spin} for $n=0$ with the
identification $\Lambda_0(t)\to\lambda(t)$ and $W_0(t) \to W(t)$.
Thus, solving the system of equations with the coupling parameter
$\Lambda_n(t)$, we can reproduce the desired population  inversion with
the cavity field initially containing $n$ photons. 
Thus, it is important to note that we have to assume the cavity mode is 
prepared with $n$ photons to reproduce $W_n(t)$ and apply the coupling
$\Lambda_n(t)$.
As an example, let us consider an arbitrary scenario where we have a
cavity containing $n$ photons.
Suppose that we want that the atom's probability of being in the ground
state to never be greater than its probability of being in the excited
state. 
In this case, the population inversion can be expressed as
\begin{equation}
  \label{eq:pop_inversion_cos_squared}
    W_n (t) = \cos^2 \left(\frac{\Omega_{n}t}{2}\right).
\end{equation}
Using this population inversion as input in Eq.
\eqref{eq:generalized_inverse_problem_coupling}, we obtain the coupling
parameter given by
\begin{equation}
  \label{eq:lambda_n_cos_squared}
  \Lambda_n (t) =
  \frac{\lambda_0 \sin \left(\Omega_{n}t\right)}
  {2 \sqrt{1-\cos ^4\left(\Omega_{n} t/2\right)}},
\end{equation}
whose behavior and population inversion it causes are represented in Fig.
\ref{fig:fig4}.

Now, for a distribution of number states, the population inversion takes
the form \cite{Joshi1993,Dasgupta1999}
\begin{equation}
\label{eq:pop_inversion_superposition}
W_\gamma(t)=\sum_{n=0}^{\infty} P_n^\gamma W_{n}(t),
\end{equation}
where $P_n^\gamma$ is a probability distribution, defined by the initial
state of the cavity mode, and $\gamma$ represents the specific
distribution of states used.
For instance, for an initial coherent state,
$P_n^\gamma \to P_n^\alpha = \exp(-|\alpha|^2) |\alpha|^{2n}/n!$, and
constant coupling $\lambda_0$, the population inversion is given by
$W_{\alpha}(t)$ in Eq. \eqref{eq:population_inversion_not_deformed_coherent}.

\begin{figure*}
  \centering
  \includegraphics[width=0.7\textwidth]{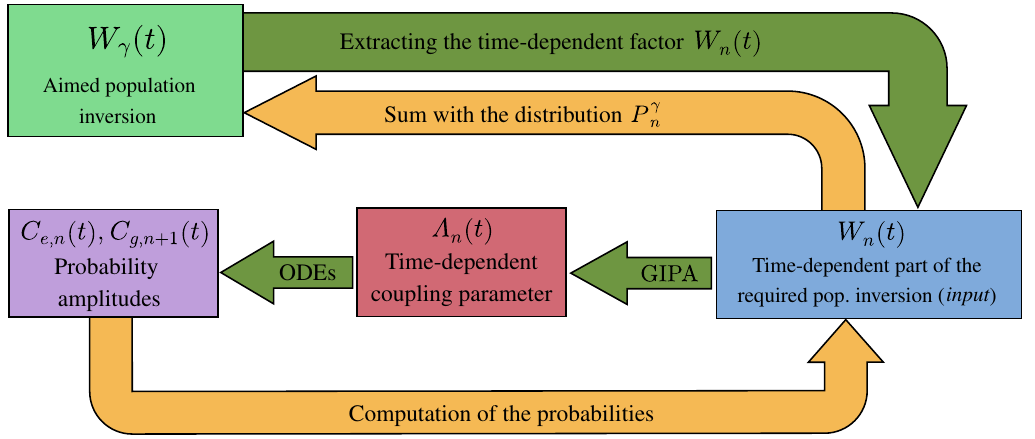}
  \caption{
    Schematic representation of the GIPA.
    We start from the top left with $W_\gamma(t)$ (green), which represents
    the population inversion we aim to reproduce.
    We extract the time-dependent part from this, $W_n(t)$ (blue).
    We apply the GIPA with $W_n(t)$ as input and obtain the
    time-dependent coupling parameter $\Lambda_n(t)$ (red).
    By solving the differential equations (ODEs) with this parameter, we
    find the coefficients $C_{e,n}(t)$ and $C_{g,n+1}(t)$ (purple).
    Upon calculating the probabilities, determining $W_n(t)$ from them,
    and summing considering the distribution $P_n^\gamma$, we retrieve
    $W_\gamma(t)$.
  }
  \label{fig:fig5}
\end{figure*}

The approach proposed here focus on $W_{n}(t)$ rather than
$W_\gamma(t)$, i.e., we  need to consider which population inversion
when summed over $n$ and multiplied by $P_n^\gamma$ yields the
population inversion related to the specific distribution of the initial
state of the cavity.
For example, if the desired population inversion is $W_{\alpha}(t)$ in Eq.
\eqref{eq:population_inversion_not_deformed_coherent}, we consider as
input $W_n(t)=\cos(\Omega_{n}t)$ and $\Lambda_n(t)=\lambda_0$.
This approach is schematically represented in Fig \ref{fig:fig5}.

\begin{figure}[b]
  \centering
  \includegraphics[width=\columnwidth]{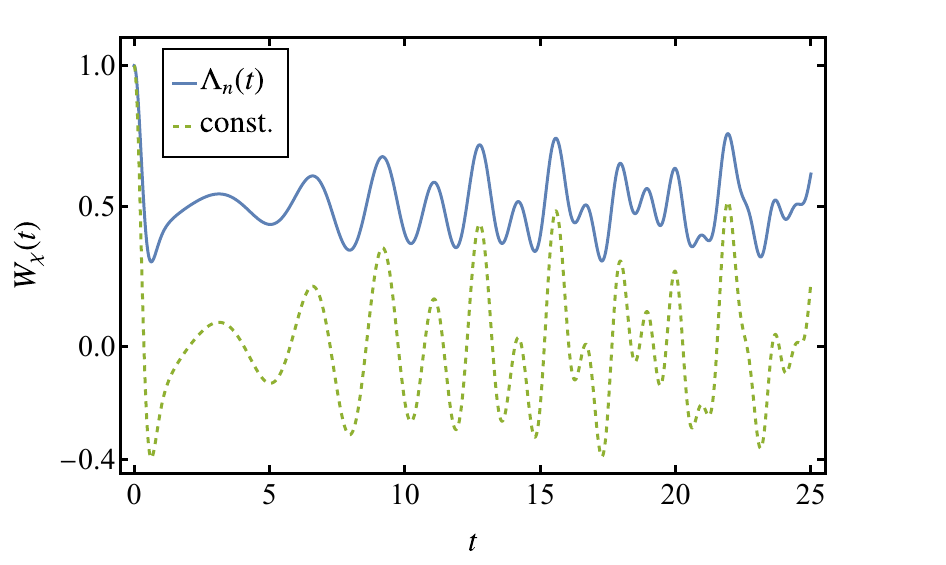}
  \caption{
    The population inversion $W_{\chi}(t)$ in
    Eq. \eqref{eq:pop_inversion_thermal_1}, both in the constant (green,
    dashed line) and time-dependent (blue, solid line) coupling
    scenarios, plotted employing
    $\lambda_0=1$, $\left\langle n\right\rangle=5$.}
  \label{fig:fig6}
\end{figure}

As a less trivial example, let us analyze the situation of an initial
thermal state of the cavity \cite{Knight1982}.
It is important to emphasize that even though the thermal state is
a mixed state, Eq. \eqref{eq:pop_inversion_superposition} is still valid
\cite{GERRY2005}, with $P_n^\gamma$ being the Bose-Einstein distribution
\cite{Knight1982}.
Let us say that the aimed population inversion is
\begin{equation}
  \label{eq:pop_inversion_thermal_1}
  W_{\chi}(t) =
  \sum_{n=0}^{\infty} P_n^\chi
  \cos^2 \left(\frac{\Omega_{n}t}{2}\right),
\end{equation}
with $P_n^\chi = \langle n\rangle^{n}/(\langle n\rangle + 1)^{n+1}$.
Applying the generalized IPA (GIPA) discussed above, with $W_n (t)$ in
Eq. \eqref{eq:pop_inversion_cos_squared} as input, we obtain
$\Lambda_n(t)$ in Eq. \eqref{eq:lambda_n_cos_squared}.
In this case, the population inversion's behavior is illustrated in Fig.
\ref{fig:fig6}, where we observe that the time-dependent coupling
results in a positive value for the population inversion.
Furthermore, compared to the constant coupling case, the amplitude of
the Rabi oscillations also decreases.
Different from the case where we consider a specific Fock state of the
cavity, it must be clear that the integer $n$ is not merely a prescribed
number, but we must consider its role in the summation over $n$ on the
population inversion, Eq. \eqref{eq:pop_inversion_superposition}.
This counterintuitive aspect of the coupling $\Lambda_{n}(t)$, creates
an additional difficulty since we cannot simply fix a given $n$ and
describe its contribution in the time behavior of $W_{\chi}(t)$.

Finally, let us return to the $\kappa$-deformed system.
Thus, if one seeks for the coupling parameter that generates the
population inversion $W^{\epsilon}_{\alpha}(t)$ in
Eq. \eqref{eq:deformed_pop_inversion}, with an initial coherent state in
the cavity, we must employ the GIPA, with the argument of the summation
found in this population inversion, specifically the part in brackets.
In this situation, the time-dependent coupling generating the corresponding
$W_\alpha^\epsilon(t)$ is given by
\begin{widetext}
  \begin{equation} \label{eq:coupling_gipa_deformed}
    \Lambda_n^{\epsilon}(t)=
    \lambda_{0}
    \frac{
      \sin\left(\Omega_{n}\right)
      - \epsilon \expval{n}
      \left[
        \sin\left(\Omega_{n}\right)
        -\sqrt{\frac{n+3}{n+1}}\sin\left(\Omega_{n+2}t\right)
      \right]
    }
    {
      \sqrt{1-\cos^{2}\left(\Omega_{n}\right)
        -2\epsilon\expval{n}
        \left[\cos\left(\Omega_{n}\right)
          \cos\left(\Omega_{n+2}t\right)
          -\cos^{2}\left(\Omega_{n}\right)
        \right]
      }
    }.
  \end{equation}
\end{widetext}
Thus, when the above coupling parameter is applied a system in an initial
state $\ket{\Psi(0)}=\ket{e,\alpha}$, we obtain the same effects as the
ones caused by $\kappa$-deformation on the population inversion.

Despite the previously mentioned complication regarding the $n$ in the
coupling expression, we can consider a context where this approach is
valuable.
For instance, let us consider a scenario where a JC system is placed in
a non-flat spacetime, with the aforementioned initial state.
If the effects of the curvature on the population inversion could be
translated into a time-dependent coupling similar to
$\Lambda_n^{\epsilon}(t)$, in the same way that the movement of the atom
within the cavity can be translated in a time-dependent coupling
\cite{Schlicher1989}, the relationship between $\kappa$-deformation and
quantum gravity maybe become clearer.

\section{Conclusions}
\label{sec:conc}

The IPA allows us to obtain a time-dependent coupling parameter from a
prescribed population inversion.
In this work, we explored the scope of this method and have shown how we
can reproduce the collapses and revivals of the Rabi oscillations in a
cavity mode initially in the vacuum, taking into account a coupling
parameter that exhibits a behavior replicating the form of the
population inversion.

Furthermore, we described the population inversion
$W_{\alpha}^{\epsilon}(t)$ of the $\kappa$-JC for an initial coherent
state on the cavity and, subsequently, apply the IPA procedure.
We derived a time-dependent coupling parameter
$\lambda_{\alpha}^{\epsilon}(t)$, which differs subtly from the
undeformed coupling parameter $\lambda_{\alpha}(t)$.
For deeper insights, we extended the IPA for a distribution of number
states.
We aimed to unveil the specific coupling parameter that reproduces the
effects of $\kappa$-deformation for an initial superposition of cavity
states.
We derived an expression for the coupling $\Lambda_n^\epsilon(t)$,
which has no trivial physical interpretation concerning its dependence
on $n$, a quantity having no individual measurement because it appears
to be summed up in the population inversion calculation.

The reproduction of quantum deformation effects through a time-dependent
coupling parameter in the undeformed JC suggests that, eventually, it
will be possible to perform simulations investigating other aspects of
the $\kappa$-JC.
The observation of macroscopic Rabi oscillations in Josephson Junction
qubits \cite{Yu2002,Il’ichev2003} indicates that, eventually, it may be
possible to achieve precise control of the coupling parameter in systems
correlated to the JC.
In this sense, the GIPA presented here is noteworthy, as controlling the
transitions of the two-level system play a significant role in quantum
control and quantum computing applications
\cite{Nielsen2010,Hernandez-Sanchez2024}.
Moreover, considering the relationship between the JC and relativistic
systems \cite{Rozmej1999,BERMUDEZ2007,Bermudez2008}, it is reasonable to
suggest that the procedure presented here could lead to fruitful
simulations of quantum deformation effects in these relativistic systems
as well.

\section*{Acknowledgments}
The authors thank Prof. Jonas Larson, Dr. Danilo Cius, Dr. Gustavo M.
Uhdre and MSc. Alison A. Silva, for helpful discussions.
This work was partially financed by the Coordenação de Aperfeiçoamento
de Pessoal de Nível Superior (CAPES, Finance Code 001).
It was also supported by the Conselho Nacional de Desenvolvimento
Científico e Tecnológico (CNPq) and Instituto Nacional de Ciência e
Tecnologia de Informação Quântica (INCT-IQ).
F.M.A. acknowledges CNPq Grant No. 313124/2023-0.

\bibliographystyle{apsrev4-2}
\input{jcgipa-rev.bbl}

\end{document}

%% file: jcgipa-rev.bbl
%